# Synergies between SALT and Herschel, Euclid & the SKA: strong gravitational lensing & galaxy evolution

**Stephen Serjeant[1]**

*Dept. Of Physical Sciences, The Open University*
*Milton Keynes, MK7 6AA, UK*
*E-mail:* `stephen.serjeant@open.ac.uk`

Gravitational lensing has seen a surge of interest in the past few years. The handful of strong lensing systems known in the year 2000 has now been replaced with hundreds, thanks to innovative multi-wavelength selection, and there is an imminent prospect of thousands of lenses from Herschel and other sub-millimetre surveys. Euclid and the Square Kilometre Array promise tens or even hundreds of thousands. Gravitational lensing is one of the very few probes capable of mapping dark matter halo distributions. Lensing also provides independent cosmological parameter estimates and enables the study of galaxy populations that are otherwise too faint for detailed study. SALT is extremely well placed to have an enormous impact with follow-up observations of foreground lenses and background sources from e.g. Herschel, the South Pole Telescope, the Atacama Cosmology Telescope, Euclid and the Square Kilometre Array. This paper reviews the prospects for high-impact SALT science and the many constraints of galaxy evolution that can result.

[1]Speaker

# 1. Introduction: the past of gravitational lensing

## 1.1 A quick lensing primer

The gravitational deflection of light in general relativity was first correctly predicted a century ago this year [1]. Figure 1 shows schematically how a background source can be gravitational lensed by a foreground deflector. Lensing depends only on the foreground matter distribution, and is independent of whether the foreground matter is luminous or dark, smooth or clumpy, in equilibrium or not.

Figure 2 shows a simulated gravitationally lensed image. Apart from not being an astronomical image, the simulation demonstrates many important features of gravitational lensing in astronomy. Firstly, lensing conserves surface brightness (a consequence of lensing conserving the phase space density of photons). Therefore, flux magnification is always accompanied by angular magnification. Secondly, multiple images are possible, such as the two mouths in the image, and as also indicated in Figure 1. Finally, some parts of a background object may be more magnified than other parts, an effect known as differential magnification (e.g. [2]). Gravitational lensing is purely geometrical, in that the photon paths do not depend on the photon wavelengths, but if the background object has a foreground magnification gradient, the observed colours of a lensed background object may not necessarily be representative of how it would appear without lensing. This change in a background object's colour can happen without any dust obscuration in the foreground lens, though extinction may also itself cause colour changes.

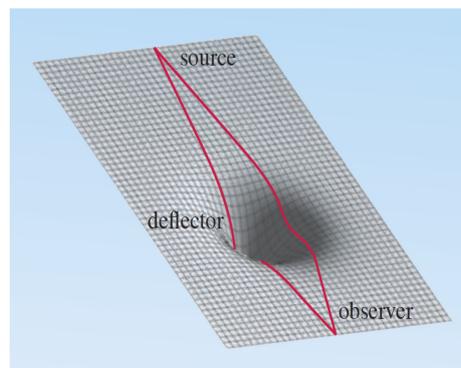

Figure 1: Schematic representation of strong gravitational lensing, showing the deflected light paths from a background source to the observer. Note that despite the term `lensing', the images are not focussed in general. Figure taken from [3].

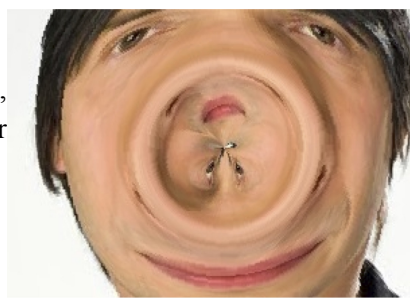

Figure 2: A background (television) star, gravitationally lensed by a transparent singular isothermal sphere with a small shear component.

### 1.2 A quick run through previous gravitational lensing surveys

Historically, gravitational lenses have most commonly been discovered through observations of candidate foreground objects. For example, Hubble Space Telescope (HST) imaging of galaxy clusters has revealed many gravitational lens arcs, and lensing has been discovered through HST follow-up imaging of Sloan Digital Sky Survey (SDSS) objects that appear to have emission line spectra differing from their absorption lines.

The advantage of selecting lenses in this way is that it has resulted in the largest published compilations of lensing systems to date. For example, the Sloan Lens ACS Survey (SLACS) has produced an impressive compilation of 131 lensing systems [4], and the Baryon Oscillation Sky Survey Emission Line Lens Survey (BELLS) has detected dozens of lenses and candidate lenses (e.g. [5]). The SLACS project yielded the surprise discovery of a double Einstein ring or "jackpot" lens [6], with which one can constrain the dark matter density profile. Such systems can only be discovered in large lensing compilations. The combination of weak and strong lensing in the large SLACS sample also statistically constrains dark matter density profiles [7]. Meanwhile, cluster lensing has a long pedigree in making high-redshift populations accessible, most recently in the HST Frontier Fields project.

The principal disadvantage of discovering gravitational lens systems through studies of plausible foreground deflectors is that it provides an unrepresentative sample of the objects providing the optical depth to gravitational lensing. Largely for this reason, samples compiled in this way are difficult to use for cosmography (e.g. [8], [9]), despite the enormous promise of gravitational lensing for constraining cosmological parameters (e.g. [10]).

### 2. The present status of gravitational lensing: a selective review

### 2.1 Finding gravitational lenses with submm and mm-wave surveys

Submm-wave surveys have been strongly diagnostic of galaxy evolution models partly because of the steep Rayleigh-Jeans tail of galaxies' dust emission in the submm, which leads to a now-famous negative K-correction. Because of this negative K-correction, a galaxy at redshift $z\sim1$ would be equally bright at the same observed submm wavelength at $z\sim5$ (e.g. [11]). The Herschel Space Observatory, launched in 2009, made over a thousand-fold improvement in submm survey mapping speed, revolutionizing extragalactic submm astronomy. In the course of its four year lifetime, Herschel has mapped approximately 10% of the sky in the submm, mainly as part of large legacy survey projects, such as:

1. The Herschel Astrophysical Terahertz Large Area Survey (H-ATLAS, [12]): the largest Herschel extragalactic survey with a catalogue of ~300,000 sources, and covering a total of 550 deg$^2$ in the North Galactic Pole (NGP), South Galactic Pole (SGP) and the three equatorial Galaxy And Mass Assembly (GAMA) survey fields. Both the GAMA and SGP fields are covered by the Visible and Infrared Survey Telescope for Astronomy (VISTA) Kilo-degree Infrared Galaxy survey (VIKING, e.g. [13]).
2. The Herschel Multi-Tiered Extragalactic Survey (HerMES, [14]), which mapped ~100 deg$^2$ to a range of depths, from shallow tiers to ultra-deep confusion limited surveys, covering the famous degree-scale multi-wavelength extragalactic survey fields (e.g.

      GOODS, XMM-LSS, ELAIS/SWIRE fields, Groth Strip, AKARI SEP, and gravitational lensing clusters).
3. The HerMES Large Mode Survey (HELMS) covered 80 deg$^2$ in the Stripe 82 SDSS field [14].
4. The Herschel Stripe 82 Survey (HerS, [15]) extended the HELMS coverage in the Stripe 82 field by a further 280 deg$^2$.

In parallel, the Atacama Cosmology Telescope (ACT, e.g. [16]) has been mapping the Southern sky at wavelengths of 1.4mm and 2.1mm. Although primarily a CMB experiment, the facility also detects infrared luminous galaxies as well as e.g. Sunyaev-Zeldovich clusters (e.g. [17]). Over the next five years, ACT will match the source density of the H-ATLAS 500μm survey over most of the Southern sky.

    Prior to Herschel's launch, several groups predicted that steep submm source counts, together with the high redshifts of submm-selected galaxies (caused by the negative K-corrections) lead to a strong gravitational magnification bias (e.g. [18], [19], [20], [21]). Shallow, wide-area submm surveys could therefore be exploited as an extremely efficient means of selecting strong lensing events, once the obvious contaminant populations of nearby galaxies and blazars are removed. Traditional methods for finding lenses include searching for high-z emission lines superimposed on early-type spectra, used by the SLACS and BELLS surveys. Unlike these projects, submm selection discovers lensing purely from its magnification and irrespective of the nature of the magnifier (e.g. early-type galaxies, spirals, groups, clusters). Moreover, the higher redshift background sources (again from the negative K-corrections) can yield much higher redshift foreground lenses than e.g. optical lensing surveys such as SLACS, making submm selection capable of probing the evolution of dark matter halos at much higher redshifts.

    In one of the major early milestones of Herschel, the lensing prediction was spectacularly confirmed in the H-ATLAS project by [22]: of the first 11 sources with S(500μm)>100mJy, 5 were obvious local galaxies, one an obvious blazar, and the remaining 5 all strong lensing systems. After removing the obvious contaminants the selection efficiency for strong lensing systems in bright submm surveys approaches an astonishing ~100%. Simple selection on bright submm fluxes yields ~100 lenses in H-ATLAS, but with more input information this increases dramatically. With submm photo-z accurate to ~30% and calibrated from our CO redshifts in > 20 H-ATLAS galaxies, we can now be reasonably confident of identifying high source redshifts, and use the steepness of the bright luminosity function as the source of magnification bias ([23], [24]). [24] showed that many of the sources with submm colors indicative of high redshifts, have high likelihood associations with K-band sources from the VIKING near-IR survey (< 10% probability by chance) consistent with much lower redshifts. Our models predict that of the sources selected by (a) submm colours indicative of high redshift plus (b) apparent foreground K-band ID, ~72 % are being strongly lensed (magnification factor μ > 2; [24]). Furthermore, the discovery of the submm-selected gravitational lens populations has stimulated other creative infrared lens detection methods. For example, [25] identified a population of submm-excess galaxies in spectral energy distribution fits to the HerMES galaxies, exploiting the comprehensive multi-wavelength data sets in the HerMES fields. These submm-excess galaxies appear to be a background submm-bright galaxy being lensed by another foreground IR-luminous galaxy.

These gravitational lens discoveries have already had an enormous impact. For example, most are well-placed for ALMA follow-ups (e.g. [26]). Breakthrough ALMA early science data proved that CO emission line redshifts of the background sources are extremely easy to obtain (e.g. 74 to date from Herschel surveys alone), as well as clearly demonstrating the advantages for exploiting high angular magnification of the background galaxies from gravitational lensing [27]. This has been followed by a spectacular ALMA image of the z=3.042 H-ATLAS lensing system SDP81 [28], from which CO kinematics reveals an unstable disk on 50−700kpc scales and a star formation efficiency ~65× local values, and identifying five giant star forming regions strikingly offset from local Larson relations possibly indicating a high pressure ISM (e.g. [29], [30], [31]).

Gravitational lensing also makes other emission line diagnostics accessible in submm-selected galaxies (e.g. [32]), but there remains the problem of differential magnification (e.g. [2], [33]). However, [34] came up with an ingenious solution: although the $H_2O$ lines detected by the authors cannot be spatially resolved in lensed submm galaxies, their line profiles are strikingly similar to selected CO transitions that can be resolved spatially, so the water emission can be corrected for lensing.

Finally, large samples of lenses selected purely on the basis of their magnification make cosmographic constraints possible (for a known population of foreground lenses, e.g. [10]) or constraints on dark matter halo populations (for a given cosmology, e.g. [35]).

## 2.2 Following up foreground redshifts: a natural niche for SALT

While the background source redshifts are easily obtained through submm/mm-wave spectra, the foreground gravitational lenses are unexpectedly much more challenging. This is because they are frequently early-type galaxies requiring 10m-class absorption line redshifts, as well as being at higher redshifts than all previous gravitational lens surveys. There is therefore a bottleneck in foreground lens redshifts, which are nonetheless critical for much of the gravitational lens modelling and interpretation.

I believe this is an ideal niche for SALT to occupy. The submm-selected gravitational lenses are distributed all over the sky, so they would be ideal for the queue-scheduled operation of SALT. Redshifts would also not require photometric conditions. Such a project would strengthen the international impact of SALT as a facility, giving it a central role in a high-profile international program. The new population of gravitational lenses is arguably the most significant extragalactic discovery on Herschel, and is a major focus of activity in many Herschel legacy survey consortia as well as the ACT. A redshift measurement programme on SALT would generate a well-cited legacy data product that would place SALT at the center of a surge of interest in lensing, and would have a long term legacy of enabling a very wide range of further science, e.g.:

- Dark matter halo substructure evolution: this is a key test of structure formation models (e.g. [36]), but there are difficulties in separating the foreground lens light from that of the background galaxy, and in distinguishing dark matter substructure from differential dust reddening in the lens. Here the background galaxy is mainly only detectable in the submm/mm-wave completely avoiding these systematic errors (e.g. [31]). Moreover the foreground lenses are detectable to much higher redshifts than purely optical lensing surveys, probing substructure evolution at much higher redshifts.

- The Initial Mass Function (IMF): by combining mass measurements from lensing with population synthesis models of the lenses (using SALT spectra & follow-up spectra) and velocity dispersions, the IMF can be constrained in lenses (e.g. [37]), already suggesting Salpeter IMFs in z~0.1−0.2 spiral bulges but Chabrier IMFs in their disks (e.g. [38]). With a large sample of lenses extending to much higher redshifts, a lens redshift programme would enable e.g. 10m-class studies of the high-z IMF.
- Resolving the background sources in rare high-magnification events (e.g. $\mu > \sim 100$). These rare systems are only discoverable in large lensing surveys. Lensing conserves surface brightness so high flux magnification is also high angular magnification. E-ELT, SKA and ALMA follow-ups would then resolve star formation on ~10pc scales in a sample of galaxies distributed throughout almost the entire Hubble volume, helping determine what causes the dramatic evolution in the demographics of star forming galaxies (e.g. [39], et seq.).
- Discovery of rare 'jackpot' lenses, in which three co-aligned galaxies lead to double Einstein rings (e.g. [6]). The two lines of sight probe the density profile in the nearest lens, and at least one background galaxy will have an easily accessible redshift from CO. Up to 1 in 40 systems are 'jackpot' lenses, providing an independent geometrical measure of the dark energy equation of state.

Since this keynote presentation was given at the SALT conference in June 2015, the Herschel extragalactic survey consortia have combined efforts on a long-term foreground lens redshift programme on SALT (PI Stephen Serjeant, Project Scientist Lucia Marchetti), extending previous SALT lens redshift programmes (PI Lerothodi Leeuw). The project opted for long-term status rather than a Partnership Proposal because constituent consortia were not able to guarantee to release targets to any and all people at any SALT institute (because this would effectively constitute a public data release), but cordially invites people at SALT institutions with lensing expertise to apply to be co-Is.

## 3. A golden future: Euclid and the Square Kilometre Array

The forthcoming 1.2m diffraction-limited Euclid space telescope will perform an imaging and slitless spectroscopy survey of around half the sky at optical and near-infrared wavelengths in order to measure the dark energy equation of state through baryon wiggles and weak lensing [40]. A byproduct of this survey will be up to $\sim 10^5$ gravitationally lensed systems that can in principle be discovered through their morphologies [40]. While it is easy in principle to create a highly complete catalogue of strong lensing systems from the Euclid imaging, it is not at all obvious how to make the sample reliable. False positives could for example be caused by arc-like tidal tails in merging galaxies, or by random lensing-like locations of HII regions in a galaxy. The size of the Euclid catalogue makes visual inspection of all the candidates a formidable task. For this reason, arc-finder algorithms are being developed for multi-wavelength imaging data (e.g. [41]). There is also a citizen science mass participation experiment to discover gravitational lenses, Spacewarps [42].

However, a subset of Euclid lensing systems can be extracted highly reliably. The Hα luminosities of star forming galaxies tend to plateau as star formation rate is increased, because of increased dust obscuration, so it is reasonable to expect the bright end of the Hα galaxy luminosity function to be steeper than the bright end of the far-infrared galaxy luminosity function. Hα observations already favour a Schechter function (e.g. [43]). The brightest Hα emitters should therefore be prone to magnification bias in the same way that bright submm galaxies are often lensed (e.g. [22]). [44] showed that by selecting Hα-emitting galaxies at $>12L_*$, Euclid will provide a sample of ~1000−3000 strongly lensed galaxies with 97%−99% reliability, i.e. less than 3% unlensed contaminants. This is shown in Figure 3. The range in the prediction comes from the range in assumed maximum magnifications of $\mu_{max}$=10−30 (corresponding to source physical sizes of 1-10$h^{-1}$kpc), i.e. more lensing systems are discoverable with larger maximum magnifications. These galaxies will already have the background redshift known from the Hα line, so would be excellent targets for 10m-class spectroscopy for the foreground lens with facilities such as SALT, as well as easy targets for Hα and Paschen α imaging by the Thirty Metre Telescope or the European Extremely Large Telescope.

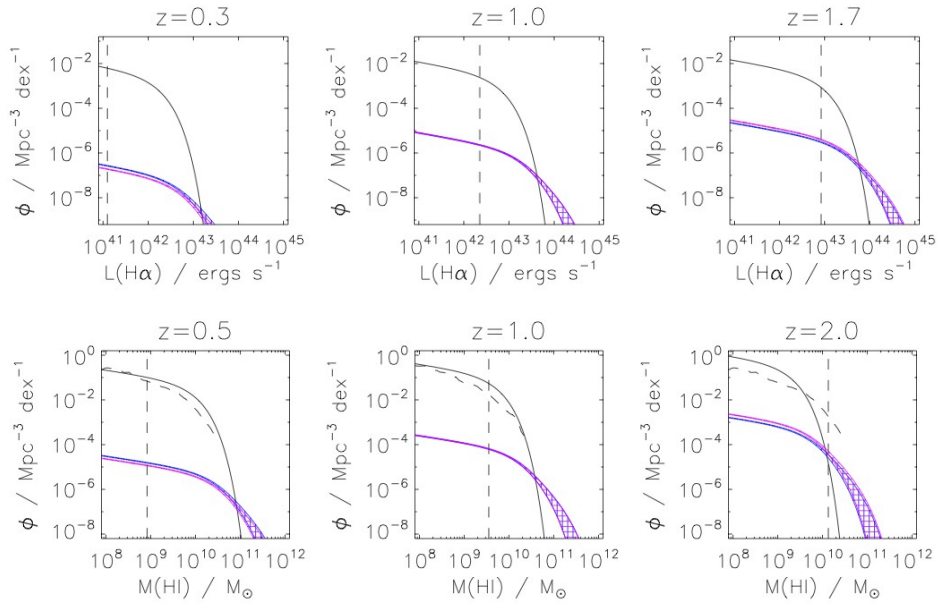

Figure 3: The top panels show the Hα luminosity function of [43] at three redshifts (black line), with the Euclid wide survey limit shown as a vertical dashed line. The lensed population is shown in the hatched areas, spanning maximum magnifications of 10 to 30. The blue hatched area shows a non-evolving lens population described in [44], while the magenta hatched area is the population model of [19] and [20]. The lower panels show the SKA predictions discussed in the text, with the same symbols as the upper panels, and the SKA phase 2 dark energy survey limit shown as a vertical dashed line. The curved dashed line is the semi-analytic prediction of [46]. Figure adapted from [44].

The situation is better still with the advent of the Square Kilometre Array, as shown in Figure 3. The neutral hydrogen mass function is expected to have a steep decline at the high mass end, so surveys of HI-emitting galaxies with the SKA and pathfinder/precursor facilities should be able to select gravitational lenses very efficiently. For example, [44] showed that making a conservative assumption of an unevolving HI mass function, a fiducial 1-year dark energy survey with the full SKA would detect between 5800 and 14,000 reliable strong gravitational lens systems at >12M$_*$. As before, the range represents maximum magnifications of $\mu_{max}$=10−30. If one instead assumes an evolving HI mass function, the effect of magnification bias is enormously increased. With the mass function from [45], the lensing prediction ranges from 7400 with an unevolving lens population and a maximum magnification of 10, to a spectacular 190,000 lensing systems with the lens population of [19] and [20].

The SKA lenses, Euclid lenses and submm/mm-wave lenses all share the advantage that the lens discovery is made purely on the basis of magnification, and irrespective of the nature of the deflectors. These are therefore ideal ways for example to probe the lensing optical depth of the Universe and to discover lenses at higher redshifts than would otherwise be accessible. SALT will once again be a natural choice for lens redshift determination.

## 4. Acknowledgements

SS is very grateful to the conference organisers and to the UK SALT consortium for financial support, to the Science and Technology Facilities Council for support under grant ST/J001597/1, and to many colleagues in the Euclid Strong Lensing and Galaxy Evolution working groups for stimulating discussions.

## References

[1] A. Einstein, 1915, *Erklarung der Perihelionbewegung der Merkur aus der allgemeinen Relativitatstheorie*, Sitzungberichte, Preussiche Akademie der Wissenschaften, **47**, 831

[2] S. Serjeant, 2012, *Strong biases in infrared-selected gravitational lenses*, MNRAS, **424**, 2429 [arXiv:1203.2647]

[3] S. Serjeant, *Observational Cosmology*, CUP, Cambridge, 2010

[4] A.S. Bolton, et al., *The Sloan Lens ACS Survey. V. The Full ACS Strong-Lens Sample*, ApJ, **682**, 964

[5] J.R. Brownstein, et al., *The BOSS Emission-Line Lens Survey (BELLS). I. A Large Spectroscopically Selected Sample of Lens Galaxies at Redshift ~0.5*, ApJ, **744**, 41

[6] R. Gavazzi, et al., *The Sloan Lens ACS Survey. VI. Discovery and Analysis of a Double Einstein Ring*, ApJ, **677**, 1046

[7] R. Gavazzi, et al., *The Sloan Lens ACS Survey. IV. The Mass Density Profile of Early-Type Galaxies out to 100 Effective Radii*, ApJ, **667**, 176

[8] P. Schneider, C.S. Kochanek, J. Wambsganss, Saas-Fee Advanced Course 33, 2006

[9] T. Treu, *Strong lensing by galaxies*, ARA&A **48**, 87

[10] J. Short, et al., [arXiv: 1206.4919]

[11] A.W. Blain, M.S. Longair, *Observing strategies for blank-field surveys in the submillimetre waveband*, MNRAS, **279**, 847

[12] S.A. Eales, et al., *The Herschel ATLAS*, PASP, **122**, 499

[13] S. Fleuren, et al., *Herschel-ATLAS: VISTA VIKING near-infrared counterparts in the Phase 1 GAMA 9-h data*, MNRAS, **423**, 2407

[14] S. Oliver, et al., *HerMES: SPIRE galaxy number counts at 250, 350, and 500 μm*, A&A **518**, L21

[15] M.P. Viero, et al., *The Herschel Stripe 82 Survey (HerS): Maps and Early Catalog*, ApJS, **210**, 22

[16] D.S. Swetz, et al., *Overview of the Atacama Cosmology Telescope: Receiver, Instrumentation, and Telescope Systems*, ApJS **194**, 41

[17] D. Marsden, et al., *The Atacama Cosmology Telescope: dusty star-forming galaxies and active galactic nuclei in the Southern survey*, MNRAS, **439**, 1556

[18] A.W. Blain, *Galaxy-galaxy gravitational lensing in the millimetre/submillimetre waveband*, MNRAS, **283**, 1340

[19] F. Perrotta, et al., *Gravitational lensing of extended high-redshift sources by dark matter haloes*, MNRAS, **329**, 445

[20] F. Perrotta, et al., *Predictions for statistical properties of forming spheroidal galaxies*, MNRAS, **338**, 623

[21] M. Negrello, et al., *Astrophysical and cosmological information from large-scale submillimetre surveys of extragalactic sources*, MNRAS, **377**, 1557

[22] M. Negrello, et al., *The Detection of a Population of Submillimeter-Bright, Strongly Lensed Galaxies*, Science, **330**, 800

[23] A. Lapi, et al. *Herschel-ATLAS Galaxy Counts and High-redshift Luminosity Functions: The Formation of Massive Early-type Galaxies*, ApJ, **742**, 24

[24] J. Gonzalez-Nuevo, et al., *Herschel-ATLAS: Toward a Sample of ~1000 Strongly Lensed Galaxies*, ApJ, **749**, 65

[25] M. Rowan-Robinson, et al., *Detailed modelling of a large sample of Herschel sources in the Lockman Hole: identification of cold dust and of lensing candidates through their anomalous SEDs*, MNRAS, **445**, 3848

[26] R.S. Bussmann, et al., *HerMES: ALMA Imaging of Herschel-selected Dusty Star-forming Galaxies*, ApJ, **812**, 43

[27] J.D. Viera, et al., *Dusty starburst galaxies in the early Universe as revealed by gravitational lensing*, Nature **495**, 344

[28] C. Vlahakis, et al., *ALMA Long Baseline Observations of the Strongly Lensed Submillimeter Galaxy HATLAS J090311.6+003906 at z=3.042*, ApJL in press [arXiv:1503.02652]

[29] S. Dye, et al., *Revealing the complex nature of the strong gravitationally lensed system H-ATLAS J090311.6+003906 using ALMA*, MNRAS submitted [arXiv:1503.08720]

[30] M. Swinbank, et al., *ALMA maps the Star-Forming Regions in a Dense Gas Disk at z~3*, ApJL in press [arXiv:1505.05148]

[31] M. Rybak, et al., *ALMA imaging of SDP.81 - I. A pixelated reconstruction of the far-infrared continuum emission*, MNRAS, **451**, L40

[32] I. Valtchanov, et al., *Physical conditions of the interstellar medium of high-redshift, strongly lensed submillimetre galaxies from the Herschel-ATLAS*, MNRAS, **415**, 3473 [arXiv:1105.3924]

[33] Y.D. Hezaveh, et al., *Size Bias and Differential Lensing of Strongly Lensed, Dusty Galaxies Identified in Wide-Field Surveys*, ApJ, **761**, 20

[34] A. Omont, et al., *$H_2O$ emission in high-z ultra-luminous infrared galaxies*, A&A 551, A115

[35] S.A. Eales, *Practical cosmology with lenses*, MNRAS **446**, 3224

[36] S. Vegetti, et al., *Gravitational detection of a low-mass dark satellite galaxy at cosmological distance*, Nature **481**, 341

[37] S. Posacki, et al., *The stellar initial mass function of early-type galaxies from low to high stellar velocity dispersion: homogeneous analysis of ATLAS$^{3D}$ and Sloan Lens ACS galaxies*, MNRAS, **446**, 493

[38] A.A. Dutton, et al., *The SWELLS survey - V. A Salpeter stellar initial mass function in the bulges of massive spiral galaxies*, MNRAS **428**, 3183

[39] E. Le Floc'h, et al., *Infrared Luminosity Functions from the Chandra Deep Field-South: The Spitzer View on the History of Dusty Star Formation at 0 <~ z <~ 1*, ApJ, **632**, 169

[40] R. Laureijs, et al., *Euclid study definition report*, ESA/SRE(2011)12 [arXiv:1110.3193]

[41] R. Gavazzi, P.J. Marshall, T. Treu, A. Sonnenfeld, *RINGFINDER: Automated Detection of Galaxy-scale Gravitational Lenses in Ground-based Multi-filter Imaging Data*, ApJ, **785**, 144

[42] P.J. Marshall, et al., *Space Warps: I. Crowd-sourcing the Discovery of Gravitational Lenses*, MNRAS in press [astro-ph/1504.06148]

[43] J.E. Geach, et al., *Empirical Hα emitter count predictions for dark energy surveys*, MNRAS, **402**, 1330

[44] S. Serjeant, 2014, *Up to 100,000 Reliable Strong Gravitational Lenses in Future Dark Energy Experiments,* ApJL, **793**, 10 [arXiv:1409.0539]

[45] F.B. Abdalla, C. Blake, C., S. Rawlings, *Forecasts for dark energy measurements with future HI surveys*, MNRAS, **401**, 743

[46] C. Lagos, et al., *Cosmic evolution of the atomic and molecular gas contents of galaxies*, MNRAS, **418**, 1649